\documentclass[useAMS,usenatbib,usegraphicx]{mn2e}
\def\simlt{\lower.5ex\hbox{$\; \buildrel < \over \sim \;$}}
\def\simgt{\lower.5ex\hbox{$\; \buildrel > \over \sim \;$}}
\def\simpropto{\lower.2ex\hbox{$\; \buildrel \propto \over \sim \;$}}
\newcommand{\be}{\begin{equation}}
\newcommand{\ba}{\begin{eqnarray}}
\newcommand{\ee}{\end{equation}}
\newcommand{\ea}{\end{eqnarray}}

 \title[ FEEDBACK AND OUTFLOWS]
        {A NEW PRESCRIPTION FOR PROTOGALACTIC  FEEDBACK AND OUTFLOWS: 
WHERE HAVE ALL  THE BARYONS GONE?}
\author[ J.~Silk]{
Joseph Silk\thanks{silk@astro.ox.ac.uk}\\
Physics Department,  Denys Wilkinson Building, Keble Road, Oxford OX1 3RH}
\begin{document}

\date{Draft version \today}

\pagerange{\pageref{firstpage}--\pageref{lastpage}} \pubyear{2002}

\maketitle

\label{firstpage}

\begin{abstract}
Up to  half of the baryons inferred to once have
been  in our galaxy have not yet been
detected. Ejection would seem to provide  the most attractive explanation.
Previous numerical studies may have  underestimated the
role of winds.
I propose a solution involving a  multiphase model of the protogalactic
interstellar medium and the possibility of driving a
superwind. Simulations do not yet incorporate the small-scale physics
that, I argue,
drives mass-loading of the cold phase gas and enhances the porosity,
thereby ensuring that winds are driven at a rate that depends
primarily on the star formation rate.
 The occurrence of hypernovae, as claimed for metal-poor and 
possibly also for starburst
environments, and  the possibility of a top-heavy primordial
stellar initial mass function   are likely to have played
 important roles in 
allowing winds to prevail  in massive gas-rich starbursting
protogalaxies as well as in dwarfs.
I discuss why
such outflows are generically of order the rate of star formation
and  may have been a common occurrence in the past.

\end{abstract}
\section{Introduction}
Cosmological observation and theory have now converged sufficiently
that one can begin to make a reasonable census of the baryons in the
universe.
Four independent probes at very different epochs of the universe
yield $\Omega_b\approx 0.04.$
These are primordial nucleosynthesis $(z\sim 10^9),$ 
the cosmic microwave background
temperature fluctuation power spectrum $(z\sim 10^3),$ modelling of the Lyman alpha
forest $(z\sim 3),$ and the hot gas fraction in galaxy clusters
$(z\sim 0).$ The latter
requires knowledge of $\Omega_m$, which I take to be 0.3, as inferred
from recent large-scale structure studies (e.g., Schuecker et al. 2003; Verde et
al. 2002; Percival et al. 2002; Peacock et al. 2001).

Approximately 10  percent of the baryons are observed in spheroid stars,
4  percent  in disk stars, and 
5  percent is observed as 
hot gas in galaxy clusters (Fukugita, Hogan and Peebles 1998). 
The local Lyman alpha forest 
accounts for an additional  20 percent of the baryons 
(Penton, Shull and Stocke 2000).
% \cite{fukugita}. 
The dominant fraction is predicted from
large-scale structure simulations to be  in a diffuse
warm/hot  intergalactic medium at $10^5-10^6\rm K$ 
heated by gravitational clustering and accretion (Cen and Ostriker
1999, Dav\'e et al. 2001).
%\cite{cen,dave}.
This WHIM gas 
traces the large-scale structure  of the galaxy distribution,  and
 accounts for about 30  to 40 percent of the total baryon fraction.
There is some evidence  in support of the existence of the WHIM,
both from the diffuse x-ray background (Soltan, Freyberg and Hasinger 2002)
%\cite{soltan} 
and OVI absorption lines 
%\cite{simcoe} 
seen towards high redshift quasars (Simcoe, Sargent and Rauch 2002).

There appears to be an  apparent shortfall  
of approximately 10 to 20 percent in the global
intergalactic baryon fraction  observed relative to that  predicted initially.
These baryons  presumably 
should have cooled and 
fallen into galaxies, and, if still present,
 might be  expected  to be observable in halos,
possibly  as hot
diffuse  gas or as MACHO-like objects. Observationally, this possibility seems unlikely (see below), and a more compelling explanation for the
"missing'' baryons is that they have undergone non-gravitational heating,
perhaps as a consequence of protogalactic, starburst and/or
 AGN-induced outflows.

There is a complementary and potentially related baryonic shortfall that
is  quantitatively  as  demanding and also rather  closer to home.
Indirect evidence suggests that 
 about twice as much mass in baryons as observed today in stars  
was once present in halos in  the form of  cooled gas.
Detailed modelling of disk galaxy formation and evolution
that incorporates gas cooling,
collapse and star formation requires an initial gas fraction  
 of 10 to 20 percent (Sommer-Larsen, Gotz and Portinari 2002; Westera
et al. 2002).
%\cite{sommerlarsen, westera}. 
Perhaps not wholly coincidentally, a  gas fraction of 13-15
percent is observed in rich
galaxy clusters (Allen, Schmidt and Fabian 2002),
%\cite{balogh},
 which may be considered to be reservoirs of primordial
gas albeit contaminated by galactic outflows.
This gas fraction is not observed in galaxies, however,
despite the fact that it was assumed to
be present within the virial radius of
 the protogalaxy.

The most detailed modelling to date of our own galaxy
concludes that  there is a well-measured stellar  (and
interstellar gas) baryon fraction within the virial radius of 
approximately 6
percent (Klypin, Zhao and Somerville 2002),
%\cite{klypin}, 
whereas semi-analytic simulations of disk galaxy formation
with cosmological initial conditions 
commonly find a  mass in  hot halo  gas comparable to that in cooled
baryons (mostly stars) 
 and which  results in
x-ray emission that exceeds observational limits 
by an order of magnitude (Benson et al. 2000;
Governato et al. 2002).
%\cite{benson, governato}.
In M31, there is a similar baryon shortfall.
Apparently, independent arguments from x-ray emission and mass
modelling, combined with  galaxy formation theory,
demonstrate that all of the gas was  present initially, yet today there is
a deficiency of unaccounted baryons that is at least  30 percent and 
may be as large as 50 percent or more of the stellar mass.
Of course the baryons could still be present in the halo,
in the form of MACHOs, for which the upper limit is about 20 percent
of the halo mass (Alcock et al. 2000),  old white dwarfs,
for which the upper limit is about 5 percent of the dark halo 
(Goldman et al. 2002),
or even in the form of extremely dense, compact
gas globules (Pfenniger, Combes and Martinet  1994;
Wardle and Walker 1999).
However such schemes are contrived at best, 
and the most logical inference is that the missing galactic baryons were blown
out in the pregalactic or protogalactic phase when the galaxy was
largely gaseous. This would only make a modest contribution to the
overall cosmological  baryon fraction, of about 10 percent, enough to account for any ``missing" baryons, but may suffice, being
enriched,
to contribute to the metallicity abundance in the intergalactic medium.
I now argue that protogalactic gas ejection provides the most likely
explanation of the galactic baryon problem.

Jets, winds and photo-ionization provide the only known  means of expelling
gas by input from AGN, supernovae and OB stars,  respectively.
I will not consider early AGN feedback here, largely because the
uncertainties
are so great concerning the role of AGN in galaxy formation and
evolution,
 but I note that it is likely that AGN feedback is important for the
intracluster gas, both for understanding the entropy floor
 (Cavaliere, Lapi and Menci 2002) and the inhibition of cooling
flows (e.g. Blanton, Sarazin and McNamara 2002).

Strong supernova feedback has been incorporated into semi-analytic
treatments of 
disk galaxy modelling in order  to expel the gas (Toft et al. 2002).
%\cite{toft}.
However detailed simulations that incorporate  some of the multiphase
astrophysics
show that outflows from a typical $L_\ast$
galaxy are largely quenched by the deep gravitational potential well
(Springel and Hernquist 2002). 
%\cite{maclow, springel}.
If one accepts this result,  then one could  consider
photo-ionization or  supernova-driven
ejection of the gas before the galaxy was assembled. Winds are known
to develop more efficiently in shallow potential wells.
The gas can also be ejected by photo-ionization in very low mass objects.
However this type of explanation requires the bulk of star formation 
and heavy element enrichment to have
occurred in the pregalactic environment, when the protogalaxy still 
consists of a
collection of hierarchically merging gas-rich dwarfs,
in order to achieve ejection of a mass of gas comparable to the mass
in stars. Such outflows have indeed been invoked (Madau, Ferrara and Rees 2001)
%\cite{madau} 
to account
 for the metallicity
in the Lyman alpha forest, this hypothesis necessarily requiring enrichment via low
mass objects to avoid destructive interference by massive winds.
These much more modest supernova-driven pregalactic outflows invoke
only of order one supernova per $10^4\rm M_\odot,$ and so
can account for the observed  Lyman alpha forest
 metallicity of about half a percent of the solar value.

There are two arguments that render the drastic 
 solution of massive pregalactic
supernova-driven outflows unacceptable in the present context.
Firstly, the diffuse IGM  would be enriched at an early
epoch to an unacceptably high level, to a factor
$\sim\Omega_\ast/\Omega_b$
of solar. One sees such enrichment in clusters
but not in the high redshift IGM sampled by absorption line
measurements towards remote quasars. 
The inferred population of
 pregalactic objects is  necessarily common and nearly uniformly
distributed, and would overpollute the entire IGM.
Secondly, and even more seriously, one would
have  formed most of the stars before the massive galaxies
were
assembled, and one could not attain the observed surface brightnesses
or central  stellar densities.

What is needed is a mechanism that ejects gas in the
protogalactic environment from massive protogalaxies, and thereby
preferentially pollutes only the denser environments that are destined
to form groups and clusters. There is some evidence from
studies of Lyman break galaxies that massive
galaxies can have winds at early stages of their evolution.
 With a space density
and clustering scale  comparable to
those  of local luminous galaxies, a median star formation rate of
 $\sim 90\rm M_\odot yr^{-1},$  a median star formation age of $\sim 0.3$
Gyr, and  a  median
stellar mass of  $\sim 3\times
10^{10}\rm M_\odot $,
a significant fraction of Lyman break galaxies at $z\sim 3$ 
  are
surely the counterparts of $L_\ast$ galaxies
(Shapley et al. 2002). 
%\cite{shapley}.
 Spectral studies
of Lyman break galaxies,
using  stacked spectra (Shapley et al. 2002a),
%\cite{shapleya},
 show evidence of outflows,
which  are also independently but indirectly inferred from
 an inverse correlation with nearby Lyman alpha forest hydrogen 
and CIV metal line system absorption that
extends out to circumgalactic distances of up to $\sim 1$ Mpc or even beyond (Adelberger et
al. 2003).
%\cite{adelberger}.

Clearly, a new prescription  for galactic outflows
would be useful in confronting many of these issues.
 I propose such a  prescription  that operates
effectively regardless of the depth of the potential well of the galaxy.
The current numerical simulations lack the resolution to study the
detailed interaction of the ejecta from supernovae with the
interstellar medium. To model the physics,
I develop a simple expression for outflows that is based on the
concept of the porosity of the hot bubbles of supernova-heated ejecta
in the ambient cold interstellar medium.
I argue that porosity and  mass loading
via instabilities at the interface of the multiphase medium control
 the outflow efficiency. The resulting outflows depend  on the
 potential well depth only indirectly via the star formation
 efficiency:
the outflow rate is generically of the order of the star formation
rate  for starbursting systems, regardless of mass.

\section{Porosity and the global star formation rate}

I first review the porosity formulation previously developed in  Silk (2001).
The porosity of the interstellar medium is the product of the maximum 4-volume
of a supernova remnant-driven bubble of hot gas driven by supernovae
and limited by ambient 
pressure of the cold interstellar medium
with the   bubble formation rate   per unit volume,
that is, the supernova rate.

The filling factor $f_{hot}$ of hot gas can be expressed in terms of the
porosity $Q$ as
 \begin{eqnarray}
f_{hot}=1-e^{-Q}. 
\label{porosity}
\end{eqnarray}
The porosity, for an ambient two--phase medium described by a statistically 
uniform gas pressure
$\rho_g\sigma_g^2$, that includes  contributions from
both thermal pressure and turbulent cloud
motions,
  can be written in the form
\begin{eqnarray}Q=\frac{\dot\rho_\ast}{G^{1/2}\rho_g^{3/2}}
\left(\frac{\sigma_f}{\sigma_g}\right)^{2.7}.
\end{eqnarray}
Here $\dot\rho_\ast$ is the star formation rate per unit volume,
$\rho_g $ is the mean gas density and $\sigma_g$ is the turbulent
velocity dispersion  of the interstellar medium.
A fitting formula (Cioffi, McKee and Bertschinger 1988)
%\cite{cioffi} 
adapted to numerical simulations 
of spherical supernova remnants expanding into a uniform medium and  that
incorporates radiative cooling, has been used in deriving this expression.
I have introduced  
$\sigma_f$ as a fiducial
velocity dispersion that is proportional to $E^{1.27}_{SN} m^{-1}_{SN}
\zeta^{-0.2}_g$ and may be taken to be 17.8 km s${^{-1}}$ for an initial
supernova energy $E_{SN} =
10^{51}$ erg, where
the mass in stars formed per supernova $m_{SN}$ is  set equal to 
 200M$_{\odot}$   and $\zeta_g $, taken to be unity,
represents the heavy element abundance  relative to the solar value.
Allowance for the contribution of hypernovae
and in the  microphysics
at the interface between the hot and cold media,
discussed below, as well as
uncertainties in the
initial mass function, 
that will help enhance porosity,  means that this  estimate of
$\sigma_f$ is a lower limit.

The global star formation rate $\dot M_\ast$ is written as 
\begin{eqnarray}\dot M_\ast=\epsilon M_g\Omega,\end{eqnarray}
where $\Omega$  is the rotation rate (or inverse dynamical time for a
non-rotationally supported galaxy), $ M_g$ is the cold gas mass, and
$\epsilon$
is the star formation efficiency.
In general, $\epsilon$  is a dimensionless function that is generally taken
to be constant by semi-analytic modellers, but is given  a high value for spheroid
formation or starbursts and a low value for disk formation.
This simply follows the common folklore that disk star formation is
continuing  and
hence inefficient, whereas spheroids completed their star formation long ago,
and hence were relatively efficient.
In fact, $\epsilon$ may be dependent on galaxy mass as well as possibly other
parameters. For example
the SDSS analysis of some  100000 galaxies suggests that $\epsilon$ decreases
with decreasing galaxy mass (Kauffmann et al. 2002).
%\cite{kauf}. 

I have previously proposed an expression for $\epsilon$ that depends
explicitly on both porosity and turbulent velocity dispersion (Silk 2001).
%\cite{silk}.
 Specifically, on combining the  expression for the star formation
 rate with that for porosity,
one finds that
the porosity is given by
\begin{eqnarray}Q=\epsilon\left(\frac{\rho}{\rho_g}\right)^{1/2}\left(\frac{\sigma_f}{\sigma_g}\right)^{2.7}.\end{eqnarray}
It is useful to  define a fiducial star formation rate and efficiency when $Q=1$:
\begin{eqnarray}\epsilon_{cr}=\left(\frac{\rho_g}{\rho}\right)^{1/2}\left(\frac{\sigma_g}{\sigma_f}\right)^{2.7}.\end{eqnarray}
We expect 
 $\sigma_g$  to be of order  100 km/s for typical protospheroidal  systems.
This  should be similar in magnitude to the fiducial scale $\sigma_f$
in the protogalactic environment, if 
$\sigma_f$ were somewhat larger than the value cited above,
 which was obtained  from  adopting  a canonical
 local initial mass function and  modelling  supernova remnants as expanding spherical shells.
There are three, probably coexisting, ways by which $\sigma_f$
should be boosted to of order $\sigma_g$ even for massive protogalaxies.

Suppose firstly that  there was, for example, one hypernova
with $E_{SN}\sim 10^{53}$ erg
for every 10 Type II supernovae. 
Hypernovae are possibly the dominant type of
supernovae in metal-poor environments and in starbursts
in terms of their overall contribution
to both metallicity and to energy input into the interstellar medium.
A contribution of this order  is  suggested both for extremely metal-poor
environments and in the case of starbursts such as
M82 by observed chemical abundances and computed yields 
(Nomoto  et al. 2002), despite theoretical models
which  may prefer a lower rate (Woosley, Zhang and Heger 2002).
 The preceding expression is useful in suggesting that when
$\epsilon\sim \epsilon_{cr}$, a wind should develop,
 if $Q\sim 1$ is indeed the relevant criterion,
even in massive galaxies if the hypernova energy input once was dominant.
 Star formation is generally assumed to have been efficient in massive
early-type galaxies, and the protogalactic porosity would plausibly
have been large.
Secondly, one can plausibly imagine that the initial mass function in the early stages of galaxy evolution was top-heavy, as suggested by discussions of primordial
star formation. This could reduce $m_{SN}$ by up to
an order of magnitude. 
Finally, it is also likely that the naive analytic calculation 
of
porosity given above  
 based on spherical shell modelling significantly  underestimates the
efficiency of  driving winds because of the omission of such critical high
resolution microphysics as Rayleigh-Taylor instabilites in the
stalling supernova-driven shells.
The limited resolution of the simulations suggests that
microscopic instabilities are being neglected that may have an important
effect on the macroscopic flow. In particular, a combination
of  Rayleigh-Taylor
instabilities as the dense cool gas decelerates in response to its
interaction with the hot supernova-driven  outflows, 
and Kelvin-Helmholtz instabilities as the wind streams
by the cold gas, makes the interstellar medium highly porous
to the wind and entrains cold gas into the wind.

Two important effects of the Rayleigh-Taylor instabilities
are to punch holes through the  cold dense shell 
and to mix the cold and hot phases.
The mixing will be enhanced by the  Kelvin-Helmholtz instabilities
at the interfaces where the hot gas flows by the cold interstellar clouds.
Both the Rayleigh-Taylor and Kelvin-Helmholtz
instabilities will drive the porosity,
 entrainment and mixing  of cold and hot gas
at rates that are  outside the domain accessible to
 current galactic-scale simulations.
Individual cloud simulations, albeit in 2-D  (Klein, McKee and Colella 1994),
suggest that wind interactions drive cloud destruction by the combined
action of Rayleigh-Taylor and  Kelvin-Helmholtz instabilities,
and mass loading of the wind is a consequence of 
ensuing conductive and ablative cloud destruction (Hartquist et al. 1997).

\section{Outflow model}

I now  develop the hypothesis that high porosity, inevitably
associated with 
a  high  star formation efficiency, should suffice to drive a wind.
For dwarf galaxies, winds are inevitable as a consequence of a
starburst, since $\sigma_g\sim
\sigma_f.$
For massive galaxies, the situation is less clear since   $\sigma_g\simgt
\sigma_f.$  In general, winds are not important today from  massive  disks since 
$\epsilon$ is small. The situation may have been completely different in the
protogalactic phase when $\sigma_g$ was larger and so most likely was
 $\epsilon.$  For massive spheroids,   $\sigma_g$ is large, and one
may well need  recourse to strong feedback from hypernovae,
a top-heavy IMF and correspondingly enhanced supernova rate,
 and/or
appeal to deficiencies in the preceding
spherically symmetric analysis that surely underestimates the effects
of porosity in order to justify the 
generation of  a protogalactic wind.
Since spheroids
are the dominant stellar reservoir, this argument  suggests
 that winds played an important
chemical evolution role at the epoch of spheroid formation.
Strong feedback is required in  recent discussions of disk angular
momenta and sizes
(Sommer-Larsen, Gotz and Portinari 2002)
and  the heating of the intragroup medium (Kay, Thomas and Theuns 2002).

It is known from simulations of low mass
galaxies that the hot SN-enriched medium excavates cavities in the
interstellar medium  and blows out in a wind.
In massive galaxies, the wind loses energy as it runs into ambient
interstellar matter and is quenched by cooling losses.
If in fact the arguments in the previous section have some validity,
the effects of entrainment and porosity are considerably
under-estimated in current simulations. This means that a plausible
outcome    
is the occurrence of  quasi-adiabatic, mass-loaded outflows once the porosity
$Q$ is large.
 As long as the hot volume fraction 
and the resulting porosity are  high,
mass outflows  entrain interstellar gas  at a rate that is comparable
to the star formation rate. I will  show that the outflows are partially suppressed
if the porosity of the hot phase is low, and the outflow rate
is   then much less than 
the star formation rate.

Consider a multiphase interstellar medium. Supernovae drive bubbles of 
hot gas that are halted by ambient thermal and turbulent pressure of the
interstellar medium, including  both hot and cold phases. I do not explicitly
consider spatial and temporal correlations of the supernovae: the outflows
will be enhanced by such correlations.

I hypothesize that the outflow rate is
\begin{eqnarray}\dot M_{outflow}=\beta\dot M_\ast f_{hot},\end{eqnarray}
and is controlled by entrainment  via the wind
mass-loading factor $L$ 
and by   porosity via the filling factor of the hot phase $ f_{hot}.$
The effective load factor $\beta$ can be written as 
\begin{eqnarray}\beta=(1+L)\frac{\Delta m_{SN}}{m_{SN}},\end{eqnarray}
where $\Delta m_{SN}$ is the IMF-weighted mass ejected per supernova of
Type II and 
the amount of  mass loading can be estimated from 
the X-ray properties of the superwinds (Suchkov et al. 1996).
%\cite{suchkov}.
In particular, the metal content of winds is an especially
powerful tool,   and  it is argued that  the wind
consists predominantly  of interstellar gas  entrained in the
wind.
Most of the oxygen
in the outflows comes from the stellar ejecta in the wind.
The enrichment  observed in Chandra observations of, for example, the
dwarf starburst galaxy NGC 1569 suggests that the mass of entrained interstellar
gas is approximately 9 times the mass of stellar
ejecta in the wind
 ($L\sim 9$) 
(Martin, Kobulnicky and Heckman  2002).
%\cite{martin}.

The preceding prescription for an outflow that is not explicitly
dependent on galactic potential well is an ansatz that can be
justified, although certainly not rigorously. It lies between extreme
viewpoints to be found in the literature. For example,
Silich and Tenorio-Tagle (1998) argue that HI halos inhibit winds even
from dwarf galaxies, and Strickland and Stevens (1999)
find that multiple superbubbles precondition the interstellar medium
and help to quench winds.
However the simulations of MacLow and Ferrara (1999) find that winds
can be
driven from dwarf galaxies, but may  underestimate the
role of winds in more massive galaxies 
because of their thin disk assumption.
Indeed, chemical evolution models suggest  that even massive
ellipticals must have driven strong early  winds to account for the trend in
$[\alpha/Fe]$ increasing with mass (e.g. Matteuci 1994).
It is clear that winds must involve SN input over
a wide range of galaxy masses  simply in order to account
for the near solar iron abundance in galaxy clusters (Renzini 2002).

Consider first the limiting case of large porosity:
\begin{eqnarray}\dot M_{outflow}\approx\beta\dot M_\ast\end{eqnarray} if $f_{hot}\sim 1.$
Since $\beta\sim 1,$ we infer  that 
{\it the outflow rate is generically of order the star formation rate,}
as indeed is observed for many starburst galaxies where evidence for winds has
been obtained (Heckman 2002).
%\cite{heckman}.
The outflow rate is observed to be approximately equal to the
star formation rate in superwinds associated with starbursts.  This is
a natural consequence of a typical IMF for which $ \Delta m_{SN}\sim
10 \rm\,M_\odot$ and $m_{SN}\sim 200\rm\,M_\odot$, so that $\beta\sim
0.5$ for $L\sim 10.$
For the cases of a Salpeter, Scalo and Kennicutt IMF, respectively,
  $m_{SN}=135, 256,107 \rm\,M_\odot.$ These estimates are only for
  SNII, as appropriate to a starburst of typical duration $4\times
  10^7\rm yr.$ 
In primordial situations with $Z\simlt 10^{-4}Z_\odot,$ or even
$10^{-2}Z_\odot,$
the IMF is
likely to be top-heavy, favouring massive stars. In this situation,
$m_{SN}$ could be somewhat lower, and this would help enhance the wind efficiency.

Suppose that the wind outflow energy amounts to a fraction $f_w$ of
the supernova input energy to the interstellar medium. The
inefficiency occurs in large part because of radiative losses.
Energy balance gives
\begin{eqnarray}\dot M_{outflow}=\dot M_\ast \frac{2  E_{SN} f_w}{ m_{SN}V_{esc}^2}\end{eqnarray}
where $V_{esc}$ is the escape velocity from the galaxy and 
$f_w$ is the wind efficiency, the fraction of supernova energy
tapped by  the outflow. Now define the ejection velocity of
 supernova-enriched matter by $ V_{ej}^2=2E_{SN}/\Delta  m_{SN}.$
The load factor is then estimated by 
\begin{eqnarray}L=f_w\left(\frac{
V_{ej}}{V_{esc}}\right)^2\frac{1}{1-e^{-Q}}.
\label{load}
\end{eqnarray}
If $Q\simgt 1$, this reduces to the usual expression for starbursts:
\begin{eqnarray}L=f_w\left(\frac{ V_{ej}}{V_{esc}}\right)^2,\end{eqnarray}
and  the wind efficiency is 
\begin{eqnarray}f_w\propto L {V_{esc}}^2.\end{eqnarray}

The preceding discussion is almost certainly far too simplistic.
Mass loading will also decelerate the outflow,
and the net effect on the mass outflow rate is likely to be more
complicated than the simple linear proportionality suggested by 
equation \ref{load}.

If the porosity is low,
the wind  is reduced relative to the star formation rate according to
\begin{eqnarray}\dot M_{outflow}\approx\beta Q\dot M_\ast,\end{eqnarray} if $f_{hot}\ll1.$
In this case there is no outflow, but there will be feedback that
however is suppressed 
by a factor roughly proportional to $\sigma_g^{-0.7}:$
\begin{eqnarray}f_w\propto L {\sigma_g}^{-0.7}\left(\frac{V_{esc}}{\sigma_g}\right)^2.\end{eqnarray}

In general, the outflow rate is comparable to the  star formation rate
in starbursts: $\dot M_{outflow}\approx\beta \dot M_\ast.$
 This would imply that the 
ejected mass is expected to be on the
  order of the stellar mass for all stars formed via the starburst
  mode. Substantial  enrichment of the intergalactic medium 
therefore occurs, and 
 can be 
 estimated for the intracluster medium  from the   corresponding dilution factor:
\begin{eqnarray}\frac{M_{ej}}{M_{ej} +M_{prim}}\approx
\frac{M_\ast}{M_{ICM}}\approx 1/3,\end{eqnarray}
where $M_{ej}$ is the ejected gas in early winds, $M_\ast$ is the
stellar mass, $M_{prim}$ is the initial (unenriched) intracluster gas mass, and 
$M_{ICM}$ is the present mass of
intracluster gas.
 In deriving this estimate I have
assumed that the gas fraction in a rich cluster is approximately 15\%.
I have also assumed that  the ejecta from Type Ia supernovae is mostly
ejected in the winds.

\section{Implications}

Semi-analytic galaxy formation modelling is plagued by the problem,
common to all discussions, that low mass galaxies have strong winds but 
massive systems cool strongly and do not. This is a consequence of the
ansatz (Dekel and Silk 1986) for cold dark matter-dominated halos
which provides  the basis for feedback in all semi-analytical galaxy
formation
modelling until now.
The porosity expression proposed here is identical for all masses: the
outflow rate is of the order of the star formation rate.
Only the star formation efficiency depends on potential well depth 
($\epsilon_{cr}\simpropto \sigma_g^{2.7}$). The wind efficiency
depends only weakly on $\sigma_g.$ One now has the prospect, yet to be
implemented in actual simulations,
that both massive galaxies  including  (some of) the
Lyman break galaxies and dwarf galaxies at high redshift can undergo
strong winds. This would simultaneously alleviate both the dwarf
excess predicted and not seen at low redshift (Moore et al. 1999)
and the cooling
catastrophe at high redshift that results in the overproduction of
massive luminous galaxies that are not seen in the nearby galaxy
luminosity function (Cole et al. 2000). 
What is more, since the winds occur early and
efficiently in massive galaxies, and the star formation rate in lower mass
galaxies is predicted to be inefficient, one has the prospect of
obtaining  consistent colour-magnitude relations for disk
and elliptical
galaxies and $[\alpha/Fe]$ ratios for early-type galaxies.
These represent  important difficulties with  current models (van den Bosch
2002;
Thomas,  Maraston  and Bender 2002),
and it is clear that a change in the feedback prescription along the
lines of what is suggested here would be desirable.

In summary,
outflows can occur even from  massive galaxies since star formation
efficiency is greatest in these systems:
 $\epsilon_{cr}\propto {\sigma_g}^{2.7},$ provided that $Q\sim 1.$
It may be  necessary however
to appeal to hypernovae, a top-heavy IMF,
possible  AGN heating, or a more refined treatment of porosity than
that presented here,  in order for $\sigma_f$ to be large enough so
that $Q\sim1$ in massive galaxies.
The outflow velocity is expected to be independent of escape or rotation
velocity, as observed for  starbursts.
The proposed analytic prescription for outflows no longer
systematically sacrifices dwarf galaxies at the expense of more
massive galaxies.
All galaxies have outflows  $\dot M_{outflow}\approx\beta \dot
M_\ast,$ for interstellar turbulence velocities 
up to $\sigma_g\sim \sigma_f,$ and at least in the protogalactic
environment, this plausibly applies to all galaxies. 

 The observed enrichment of outflows suggests the entrainment
load factor is around 10. The best case is  based on Chandra 
observations of  the dwarf
starburst galaxy NGC 1569.
One would expect both the interaction of the supernova-driven hot gas
with clumps of ambient cold gas 
and Kelvin-Helmholtz instabilities at the interface with the diffuse
cold interstellar medium, once the porosity becomes large,
to enhance the
entrainment of cold gas. The colder, denser clumps of  gas should
have kinematics that reflects their dynamical entrainment.
The  maximum velocity of the entrained clouds
is at most 10 percent of the global shock velocity (Poludnenko, Frank
and Blackman 2002).
Different velocity outflows are observed for different states of the
interstellar gas. For $H\alpha$ emitting  gas, lower velocity flows are
found  than for
the x-ray emitting gas. The  lowest outflow velocities occur for the neutral
gas seen in HI.
This velocity structure is expected if    entrainment
of the cold interstellar medium is occurring.

Another consequence is enrichment and heating of the intergalactic,
and in particular the intracluster, medium.
The outflow model predicts that
 $V_w\approx V_{SN}L^{-1/2}$, where $V_{SN}$ is somewhere
between
the supernova ejection velocity  per unit mass of
gas consumed to form a supernova 
either with cooling  $E_{SN}/V_cm_{SN}$, where $V_c$  is the remnant
velocity that marks the transition from adiabatic to cooling-dominated expansion,
or without cooling $(E_{SN}/m_{SN})^{1/2}$.
This gives  $V_w\approx 300\rm km/s$ or 1 keV per particle,
similar to what is required to break cluster self-similarity
(Lloyd-Davies, Ponman and Cannon 2000; Borgani et al. 2002).
%\cite{ponman, borgani}.

The model presented here is highly simplified.
In effect, I suggest a quantitative way of empirically incorporating 
crucial pieces of microphysics that are not present in current galaxy
outflow simulations. 
An
 amount of baryons that is comparable 
to the  stellar mass currently observed in galaxies
 may have been ejected in early outflows.
Such massive early outflows could modify the dark matter halo core
profiles
(Binney, Gerhard and Silk 2001), and consequently
reduce the dark mass concentrations in massive galaxies as suggested by modeling of
gravitational lensing time delays (Kochanek 2003) as well as in low
surface brightness galaxies 
as inferred from rotation curves (de Blok and Bosma 2002).
Early massive winds might also
result in selective ejection of
 low angular momentum gas, thereby helping to alleviate the 
angular momentum problems of  galactic disks (Steinmetz and Navarro 1999;
Bullock et al. 2001).

The present model considers only a steady state, and
should   be generalized to
study the time-development of 
the load factor and the porosity. The evolving role of
hypernovae
will contribute to the evolution of feedback.
Winds are not the only outcome, as one could equally consider
galactic fountains and similar phenomena as a means of self-regulating
the rate of star formation.
There needs to be some
incorporation of threshold effects,  the clustering of supernovae,
the dispersion in
porosity and  the effects of geometry, which will result in a
minimum 
mass in stars needed before feedback becomes important, and help
 account for the 
fact that there is a  range of metallicity, gas content,
and surface brightness in dwarfs but  far more uniformity in more
massive galaxies.

\section{Acknowledgements}
I acknowledge helpful discussions with G. Bryan, J. Devriendt,
C. Martin,
P. Podsiadlowski,  A. Slyz and  J. Taylor,
and the hospitality of the KITP where this work was begun.
I also thank the referee for useful comments.

\end{document}